# Probabilistic adaptation of language comprehension for individual speakers: evidence from neural oscillations

Hanlin Wu[1], Xiaohui Rao[1], Zhenguang G. Cai[1,2,*]

[1]Department of Linguistics and Modern Languages, The Chinese University of Hong Kong, Sha Tin, Hong Kong
[2]Brain and Mind Institute, The Chinese University of Hong Kong, Sha Tin, Hong Kong

[*]Corresponding author. Department of Linguistics and Modern Languages, The Chinese University of Hong Kong, Leung Kau Kui Building, Sha Tin, N.T., Hong Kong. E-mail: zhenguangcai@cuhk.edu.hk.

## Abstract

**Listeners adapt language comprehension based on their mental representations of speakers, but how these representations are updated remains unclear. We investigated whether listeners probabilistically adapt comprehension based on the frequency of speakers making stereotype-incongruent statements. In two EEG experiments, participants heard speakers make stereotype-congruent or incongruent statements, with incongruency base rate manipulated. In Experiment 1, stereotype-incongruent statements decreased high-beta (21–30 Hz) and theta (4–6 Hz) oscillatory power in the low base rate condition but increased it in the high base rate condition. The theta effect varied with listeners' openness trait: less open-minded participants tended to show theta increases to stereotype incongruencies, while more open-minded participants tended to show theta decreases. In Experiment 2, we dissociated incongruency base rate from the target speaker by manipulating it using a non-target speaker and found that only the high-beta effect persisted. Our findings reveal two potential mechanisms: a speaker-general mechanism (indicated by high-beta oscillations) that adjusts overall expectations about hearing statements that violate social stereotypes, and a speaker-specific mechanism (indicated by theta oscillations) that updates a more detailed mental model specifically about an individual speaker. These findings provide evidence for how language processing interacts with social cognition.**

## Introduction

Spoken language comprehension is influenced by the speaker's identity (Wu and Cai 2024). In an early study, Lattner and Friederici (2003) found that when listeners hear sentences that conflict with gender stereotypes, such as a female voice saying 'I like to play *soccer*' or a male voice saying 'I like to wear *lipstick*', their brain potentials show a P600 effect at the sentence's critical final word. Van Berkum et al. (2008) examined listeners' brain responses to mismatches involving broader sociodemographic characteristics, including age and social status, in addition to gender. When participants heard sociodemographically incongruent statements, such as a child saying 'Every evening I drink some *wine* before I go to sleep', they observed an N400 effect at the critical word *wine*. Further research using functional magnetic resonance imaging localized this speaker effect to bilateral inferior frontal gyri, where increased activation suggests heightened cognitive effort in integrating speaker information with sentence meaning and world knowledge (Tesink et al. 2009). Such influence of speaker identity on language processing has been consistently demonstrated across multiple studies (van den Brink et al. 2012, Foucart et al. 2015, Pélissier and Ferragne 2022, Wu et al. 2024, Wu and Cai 2025).

### Speaker models: from population-based knowledge to individual representations

Speaker effects are often explained through the speaker model account, which posits that listeners maintain a mental model of the speaker's identity, including their biological gender, age, socioeconomic status, and region of origin. This model influences listeners' expectations and interpretation of language by integrating linguistic content with their knowledge about the speaker.

Evidence for speaker modelling in language comprehension comes from studies showing that speaker effects persist independently of acoustic information. For example, Cai et al. (2017) found that listeners' interpretation of cross-dialectally ambiguous words (like *flat* or *gas*) was influenced by their beliefs about the speaker's dialect background, even when the acoustic properties of the words were controlled for (see also King and Sumner 2015, Cai 2022). Foucart et al. (2019) showed that exposure to a speaker's accent influenced subsequent language processing even in written comprehension. The model's broad influence is further evidenced by studies showing effects of perceived speaker nationality (Niedzielski 1999), ethnicity (Staum Casasanto 2008), and age (Hay et al. 2006).

Speaker modelling extends beyond capturing general sociodemographic attributes—it also captures the unique characteristics of specific individuals. Research shows that listeners can quickly recognize familiar voices (Schweinberger et al. 1997, Beauchemin et al. 2006) and form expectations about a speaker's unique lexical choices (Barr and Keysar 2002, Metzing and Brennan 2003, Ryskin et al. 2020), syntactic preferences (Kroczek and Gunter 2021, Xu et al. 2022), communicative style (Regel et al. 2010), perspective-taking (Hanna et al. 2003, Brown-Schmidt et al. 2008, Brown-Schmidt 2012), reliability (Brothers et al. 2019), and social connections (Barr et al. 2014). These expectations are formed for specific individuals, regardless of their sociodemographic background.

### Probabilistic adaptation of speaker models: a proposal

While the influence of speaker modelling on comprehension is well-documented, a crucial unanswered question is how it is dynamically adapted through experience. When encountering a speaker who violates stereotypical expectations, how do listeners adapt their mental representations? One potential mechanism for such adaptation is probabilistic learning.

In language processing, comprehenders have been shown to rationally integrate prior expectations with linguistic input, modulating their comprehension strategies based on statistical regularities in the context. For example, Gibson et al. (2013) demonstrated that the interpretation of implausible sentences (e.g. 'The mother gave the candle the daughter'.) is sensitive to the *base rate* of their occurrence. When such sentences were rare (as in *low base rate* condition), comprehenders were likely to infer a plausible meaning by assuming a noise-corrupted input (e.g. that *to* was omitted from 'gave the candle *to* the daughter'). Conversely, when implausible sentences were frequent (as in *high base rate* condition), listeners were more likely to adopt a literal interpretation rather than inferring a more plausible alternative.

We propose that a similar probabilistic mechanism may govern the adaptation of speaker models. Listeners may track the base rate of a speaker's stereotype-incongruent statements to refine their expectations. When encountering an unfamiliar speaker, listeners may initially rely on sociodemographic stereotypes. However, through repeated interactions, they may develop more precise, individualized mental representations that reflect the unique characteristics of specific speakers.

Some prior work aligns with this proposal. For example, Grant et al. (2020) examined how listeners adapt to speakers who violate gender stereotypes. In their study, participants listened to male and female speakers producing sentences about stereotypically feminine (fashion) and masculine (sports) topics. Using ERPs, they found that N400 effects for semantic violations were larger in stereotype-incongruent conditions (e.g. male speakers making semantic errors about sports but not fashion) compared to stereotype-congruent conditions (e.g. male speakers making errors about fashion but not sports), suggesting that listeners integrate speaker characteristics with linguistic input. They also found that participants with higher sexism scores showed less adaptation to stereotype-violating speakers over time. While these findings suggest that listeners can adapt to stereotype-violating speakers, they do not directly test the probabilistic nature of this adaptation or delineate the underlying neural mechanisms.

### The current study

The current study directly investigates whether and how listeners probabilistically adapt language comprehension for individual speakers. In two EEG experiments, we manipulated the base rate (low vs. high) at which a speaker makes self-referential statements that are stereotypically incongruent with their sociodemographic profile either in terms of gender stereotype (e.g. 'This weekend I'm going to get a *manicure* and a haircut' by a male speaker) or in terms of age stereotype (e.g. 'He took my *toys* away from me' by an adult). A speaker rarely produces gender- or age-incongruent statements in the low base rate condition but frequently do so in the high base rate condition.

In Experiment 1, we tested whether listeners adapt to a single speaker's probability of making stereotype-incongruent statements. We hypothesized that if listeners track and adapt to this probability, neural markers of the stereotype-incongruency effects should be modulated by the base rate of incongruency. In Experiment 2, we further investigated whether such adaptation, if observed, reflects speaker-specific learning or a more general habituation to stereotype violations. We associated the base rate manipulation with a *non-target* speaker while measuring its effect on the processing of the *target* speaker's statements. If the adaptation is speaker-specific, the base rate manipulation of the non-target speaker should not affect the processing of the target speaker's statements. If the adaptation is a general habituation, the base rate effect should persist despite it has been dissociated with the target speaker.

To probe these adaptation mechanisms, we focused on time-frequency representations. While ERPs primarily capture phase-locked activity that typically reflect bottom-up processing, time-frequency representations can additionally reveal non-phase-locked oscillatory activity. This activity is associated with top-down, predictive processes (Chen et al. 2012, Herrmann et al. 2014) that are central to expectation-based adaptation. For example, beta-band oscillations have been linked to maintaining or changing the current cognitive state (Engel and Fries 2010), whereas theta-band oscillations are associated with memory operations (Herweg et al. 2020). Based on this, we formed the general hypothesis that the active adaptation processes involved in speaker-based comprehension adaptation would be reflected in neural oscillations. However, as no prior studies have examined this adaptation process with time-frequency representations, we did not form specific hypotheses regarding particular frequency bands or time windows. Finally, we hypothesize that this adaptation process would be modulated by individual differences in the personality trait of *openness*, which reflects a willingness to adjust existing beliefs (McCrae 1996, Flynn 2005). We predict that more open-minded individuals would show greater flexibility in adapting their speaker models in response to stereotype-incongruent statements.

## Experiment 1

### Design

In this experiment, participants listened to speakers talking about themselves in short sentences. The content of these sentences was either congruent or incongruent with the speaker's identity in terms of common social stereotypes, while the base rate of stereotype-incongruent statements varied. We adopted a 2 (Congruency: stereotype-congruent vs. stereotype-incongruent) × 2 (Base rate of incongruency: low vs. high) factorial design. Both Congruency and Base rate of incongruency were manipulated within participants and items.

Specifically, there were 60 target trials with half being stereotype-congruent and the other half being stereotype-incongruent in one block. Base rate of incongruency was manipulated by filler trials. In the low base rate condition, half of the filler trials were stereotype-congruent and the other half were stereotype-neutral (irrelevant to the speaker's gender or age); in the high base rate condition, half of the filler trials were stereotype-incongruent and the other half were stereotype-neutral.

## Participants

We recruited 30 neurologically healthy participants (15 females, 15 males; mean age = 23.90 years, *SD* = 3.18 years) who were native speakers of Mandarin Chinese. All participants provided informed consent before the experiment began. The study protocol was under the ethical standards of the Helsinki Declaration and approved by the Joint Chinese University of Hong Kong-New Territories East Cluster Clinical Research Ethics Committee.

## Materials

We constructed 120 Mandarin self-referential sentences as target sentences, along with 180 filler sentences that fell into four categories, each with 30 sentences (see Table 1 for target examples; see Table S1 for the full list, see online supplementary material for a colour version of this table). We designed target sentences following these rules: first, the sentence was always stereotypically congruent with one sociodemographic group of people (e.g. males in the gender contrast or adults in the age contrast) but incongruent with another group (e.g. females in the gender contrast or children in the age contrast); second, the stereotype (in)congruency always emerged at a critical disyllabic word; third, the critical word was always preceded by a word or words of at least three syllables (equivalent to three characters) to ensure that listeners had constructed the speaker context before encountering the critical word (research shows that listeners perceive a speaker's identity, including sociodemographic characteristics like age and gender, remarkably fast and accurately upon hearing their voice, e.g. Scharinger et al. 2011, McAleer et al. 2014, Lavan et al. 2024); forth, the critical word was always followed by at least three syllables before the sentence ended to eliminate the influence of the sentence wrap-up effect.

We conducted a stereotypicality rating on experimental sentences with 32 participants (16 females, 16 males; mean age = 22.09 years, *SD* = 0.78 years) who were not included in the EEG experiment. Participants were individually tested online using Qualtrics. To eliminate the contextual influence of the sentence content post-critical word, participants were presented with segments of experimental sentences (in text) that started from the initial word and stopped at the critical word (e.g. the sentence 'He took my *toys* away from me' would be presented as 'He took my toys…'). They were asked to rate each sentence segment on a 7-point Likert scale for their perceived stereotypicality. For gender-contrast sentences, they were asked to indicate how likely the sentence was produced by a male speaker or a female speaker (1 = extremely likely to be produced by a male speaker; 7 = extremely likely to be produced by a female speaker; male-female scale counterbalanced among participants). For age-contrast sentences, they were asked to indicate how likely the sentence was produced by a child speaker or an adult speaker (1 = extremely likely to be produced by an adult speaker; 7 = extremely likely to be produced by a child speaker, adult-child scale counterbalanced among participants). By-item analyses showed that items could be distinguished between being stereotypically male-congruent or female-congruent ($t$ (117.24) = –32.13, $P < .001$), and between being stereotypically adult-congruent or child-congruent ($t$ (98.25) = –47.28, $P < .001$).

Table 1. Examples of stimuli with English translations.

| Category | Example (English translation) |
|---|---|
| Congruent/ incongruent speaker | |
| Male/female | 在工作单位我一般都是穿西服和衬衫。<br>(I usually wear a suit and shirt at work.) |
| Female/male | 这个周末我要先去做美甲然后理发。<br>(This weekend I'm going to get a manicure and a haircut.) |
| Adult/child | 我喜欢晚上去酒吧喝酒放松。<br>(I like to go to pubs at night to drink and relax.) |
| Child/adult | 他把我的玩具抢走了。<br>(He took my toys away from me.) |

The critical word in a sentence is underscored and marked in bold.

For each sentence, we generated two versions of audio using the voices of two speakers (i.e. a male and a female speaker for the gender-contrast items; an adult and a child speaker for the age-contrast items). The sentence content was congruent with one speaker's identity but incongruent with the other in terms of biological gender (in the gender contrast) or age (in the age contrast). For gender-contrast sentences, one speaker was a male adult, and the other a female adult. For age-contrast sentences, one speaker was a male adult, and the other a male child. To ensure that the speech audio minimized differences other than the manipulated gender or age, we used text-to-speech to generate audio files, controlling for potential confounds such as volume, accent, and speech rate, which are often inevitable with human speakers. We selected standard voice profiles that were clearly and readily identifiable as belonging to the intended sociodemographic group. We included four artificial speakers in total. The duration of the critical word in target sentences was matched between the congruent and incongruent audio versions (398.03 ms vs. 400.93 ms, $t$ (120) = –0.45, $P = .651$).

## Procedure

Participants were individually tested in a soundproof booth designed for EEG signal acquisition. Each participant was tested in a block of gender contrast and a block of age contrast. The manipulation of base rate (low vs. high) was counterbalanced between blocks and among participants. For example, a participant was either tested with a low-base-rate block of gender contrast and a high-base-rate block of age contrast or a high-base-rate block of gender contrast and a low-base-rate block of age contrast. The order of blocks was counterbalanced among participants. In each block, participants listened to a speaker talking about themselves in short sentences, with each sentence constituting a trial. A block contained 120 trials, and the whole experiment had 240 trials in total. Each trial began with a fixation cross on the centre of the screen for 1000 ms. The audio was then played while the fixation cross remained on the screen until 1000 ms after the audio offset. Each trial was followed by an interval of 3600 ms. To ensure their attentive listening, participants were required to report their impressions about the speaker after each block. After the experiment, participants completed the Big Five Inventory-2 (Mandarin version, Zhang et al. 2022), of which the subscore of Openness was used in the analyses.

## EEG recording and preprocessing

The electroencephalography (EEG) was collected using 128 active sintered Ag/AgCl electrodes positioned according to an extended 10-20 system. All electrodes were referred online to the left earlobe. Signals were recorded using a g.HIamp amplifier and digitalized at a sampling rate of 1200 Hz. All electrode impedances were maintained below 30 kΩ throughout the experiment. EEG data preprocessing was performed using customized scripts and the FieldTrip toolbox (Oostenveld et al. 2011) in MATLAB. The raw EEG data were bandpass-filtered offline at 0.1–45 Hz, resampled at 500 Hz, and re-referenced to the average of the left and right earlobes (A1 and A2). Independent component analysis was applied to highpass-filtered (at 1 Hz) continuous data with dimensionality reduction to 30 components to identify and remove ocular artifacts. The data were then processed separately for the amplitude analysis and time-frequency analysis. For the amplitude analysis, the continuous data were bandpass filtered at 0.2–30 Hz, epoched from 200 ms before to 1200 ms after the onset of the critical word, and baseline-corrected by subtracting the mean amplitude from 200 to 0 ms before the critical word onset. Epochs with amplitudes exceeding ± 100 µV were considered to contain artifacts and thus excluded (5.58%). For the time-frequency analysis, the continuous data were highpass-filtered at 1 Hz and epoched from 1500 ms before to 2500 ms after the onset of the critical word. Again, epochs with amplitudes exceeding ± 100 µV were considered to contain artifacts and thus excluded (4.64%).

## Results

We analysed EEG amplitude in two time windows: 300–500 ms and 600–1000 ms post-critical word onset, corresponding to typical N400 and P600 windows respectively. Analyses focused on a region of interest of 76 central electrode sites (see Table S2 for the full list, see online supplementary material for a colour version of this table). This ROI was chosen as it is well-suited to capturing the typical scalp topographies of language-related brain responses like the N400 (Kutas and Federmeier 2011) and P600 (Osterhout and Holcomb, 1992), and was consistent with previous studies focusing speaker-content relationships (e.g. Van Berkum et al. 2008, Wu et al. 2024), while also minimizing the influence of artifacts related to movements that are more prevalent at peripheral sites. We fit linear mixed-effects (LME) models to the mean amplitudes within these windows for each trial. For all LME models, we included Participant and Item (referring to each unique sentence stimulus) as random effects to account for by-participant and by-item variability. The random-effect structure was determined using a data-driven forward model comparison procedure to find the maximal structure supported by the data (Matuschek et al. 2017; $\alpha$ = 0.2). This procedure was used only for selecting the random effects, including all possible random slopes for by-participant and by-item effects in addition to random intercepts for Participant and Item. Models with Congruency (congruent = −0.5, incongruent = 0.5) and Base rate of incongruency (low = −0.5, high = 0.5) as interacting fixed-effect predictors revealed no significant main effects or interactions in either time window. Additional models including Openness (a continuous variable) as a fixed-effect predictor interacting with Congruency and Base rate of incongruency also showed no significant effects (see Table S3 for model structures and the full results, see online supplementary material for a colour version of this table).

The time-frequency power representations were extracted using complex Morlet wavelets (Cohen 2014). Wavelet frequencies ranged from 2 to 45 Hz in steps of 1 Hz, with the number of cycles ranging from 3 to 6. Power was normalized as the relative change to the baseline from 300 to 100 ms before the critical word onset [(power - baseline)/baseline].

The subsequent analyses followed a two-step procedure. In the first step, we conducted a cluster-based permutation test (CBPT) to identify significant spatio-temporal windows of interest. To test whether the stereotype-incongruency effects differed between the base rate conditions (low vs. high), we used CBPT to find significant clusters of adjacent time-channel data points on delta (2–3 Hz), theta (4–6 Hz), alpha (7–12 Hz), low-beta (13–20 Hz), and high-beta (21–30 Hz) frequency bands, respectively. To do this, we calculated the difference between stereotype-congruent and stereotype-incongruent conditions for each data point to capture the stereotype-incongruency effect, and performed paired-sample $t$ tests between low and high base rate conditions on each data point of incongruency effect. Adjacent points with $P$ values lower than the two-tailed significant threshold of 0.025 formed clusters. For each cluster, a cluster statistic was calculated by summing the $t$ values within the cluster. To evaluate significance, condition labels of low and high base rate were permuted 1000 times using the Monte Carlo method, and the maximum cluster statistic of each permutation formed a null distribution. The observed cluster statistic was then compared with the null distribution, and a $P$ value was calculated as the proportion of permuted cluster statistics that exceeded the observed value. Clusters with $P$ values smaller than .05 were considered significant.

After identifying significant clusters, we proceeded to the second step: fitting LME models on the trial-level data to further investigate the potential cognitive processes associated with each cluster. To do this, we first extracted each cluster's trial-level data by calculating the cluster's mean power, forming a distribution. The distribution was then normalized by applying a logarithmic transformation. We added the absolute value of the minimum of the distribution and a small constant (0.001) to the distribution (as an adjustment to avoid negative or zero values) before the logarithmic transformation to ensure that it could be performed on all data points. For the LME models, the random-effect structure was determined using the same forward model comparison procedure as was used for the amplitude analysis.

As shown in Fig. 1, we identified a significant cluster associated with high-beta oscillations (220–330 ms, cluster statistic = −786.45, $P$ = .006) and a significant cluster associated with theta oscillations (320–580 ms, cluster statistic = −683.78, $P$ = .021). We fit LME models to the trial-level data of each cluster with Congruency (congruent = −0.5, incongruent = 0.5) and Base rate of incongruency (low = −0.5, high = 0.5) as interacting fixed-effect predictors (see Table S4 for model structures, see online supplementary material for a colour version of this table). Both clusters showed significant interactions between Congruency and Base rate (high-beta: $\beta$ = 0.20, $SE$ = 0.06, $t$ = 3.53, $P$ < .001; theta: $\beta$ = 0.24, $SE$ = 0.08, $t$ = 3.01, $P$ = .003). In the low base rate condition, stereotype incongruencies lead to a significant decrease in both high-beta power ($\beta$ = −0.10, $SE$ = 0.04, $t$ = −2.51, $P$ = .013) and theta power ($\beta$ = −0.12, $SE$ = 0.05, $t$ = −2.19, $P$ = .029). Conversely, in the high base rate condition, incongruencies lead to an increase in both high-beta power ($\beta$ = 0.09, $SE$ = 0.04, $t$ = 2.36, $P$ = .018) and theta power ($\beta$ = 0.12, $SE$ = 0.06, $t$ = 2.06, $P$ = .040). Models including Openness revealed a significant interaction with Congruency in the theta cluster ($\beta$ = −0.10, $SE$ = 0.04, $t$ = −2.56, $P$ = .011). Participants with openness scores below average showed a non-significant trend towards theta power increase for incongruencies ($\beta$ = 0.07, $SE$ = 0.06, $t$ = 1.15, $P$ = .253), while those with above-average openness showed a non-significant trend towards decrease ($\beta$ = −0.10, $SE$ = 0.06, $t$ = −1.71, $P$ = .089). No significant Openness-related effects were found for the high-beta cluster.

## Discussion

Experiment 1 revealed two findings in neural oscillations. First, we observed an early high-beta effect (220–330 ms after the critical word onset) and a later theta effect (320–580 ms), both being modulated by the base rate of stereotype-incongruent statements. Lower base rate led to decreased power in high-beta and theta oscillations. Second, the theta effect was further modulated by listeners' openness trait, with more open-minded individuals tending to show a theta power decrease in response to speaker incongruency, while less open individuals tended to show an increase. We did not observe significant effects in the traditional N400 or P600 components.

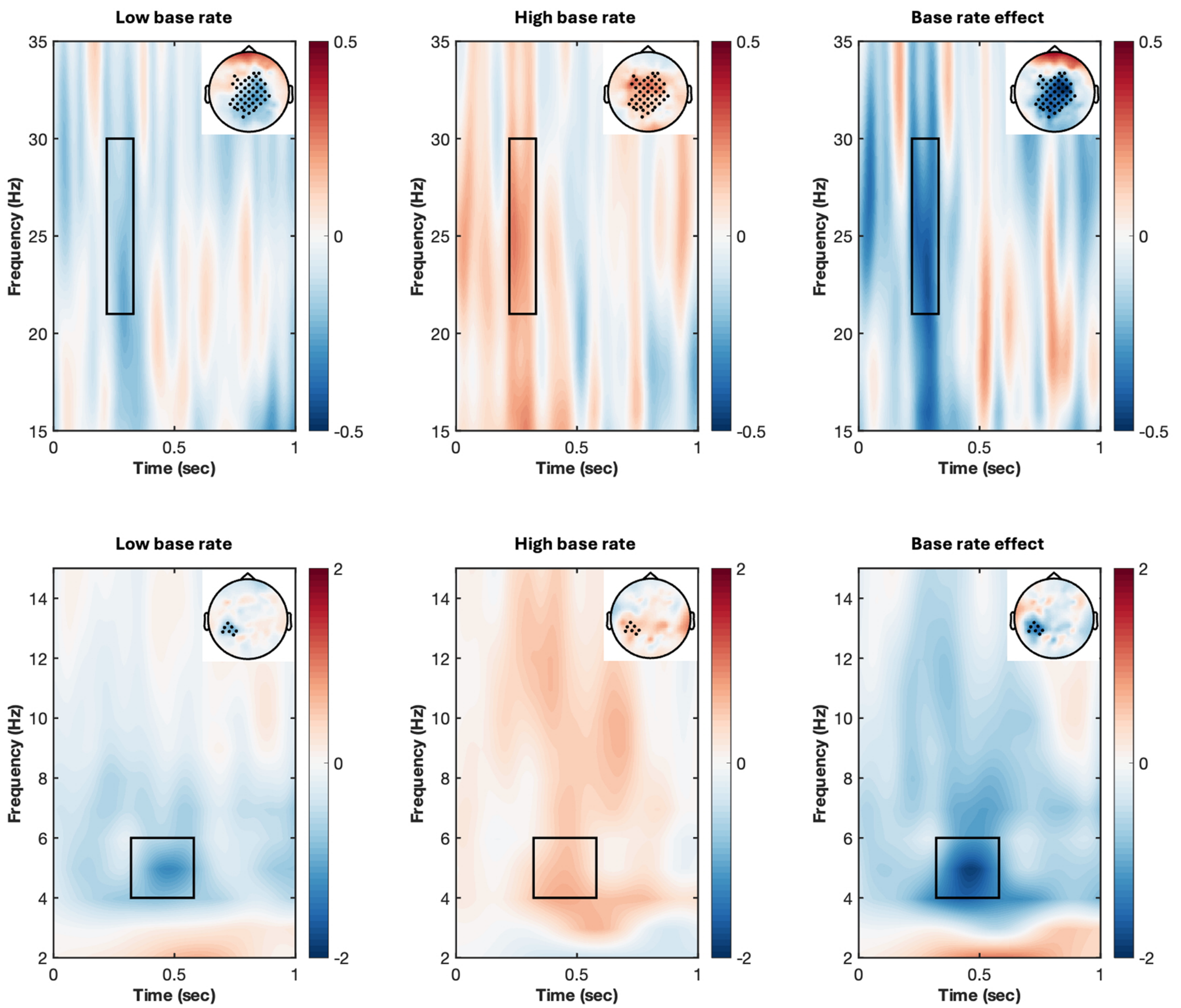

Figure 1. Time-frequency representations of stereotype-incongruency effects of the high-beta (21–30 Hz, 220–330 ms) and theta (4–6 Hz, 320–580 ms) clusters in Experiment 1.

The modulation of neural responses by Base rate suggests that listeners dynamically adjust their neural systems based on the statistical properties of the message input. This adaptation is evident in both high-beta and theta activities, indicating that the effects might occur at multiple levels of processing.

The observed adaptation could potentially arise from two distinct mechanisms. One possibility is a *speaker-general* mechanism, where frequent exposure to stereotype-incongruent statements leads to a general tolerance of stereotype incongruency. Under this account, listeners might become less sensitive to stereotype violations in general, regardless of the specific speaker violating it. Alternatively, the adaptation could reflect a *speaker-specific* mechanism, where listeners construct and update detailed mental models for individual speakers. In this case, the modulated neural responses would result from listeners learning about the speaker's unique characteristics and adjusting their expectations specifically for the target speaker. To differentiate between these possibilities, we need to dissociate speaker-specific learning from general exposure effects, which is the goal of Experiment 2.

## Experiment 2

The experimental design, materials, and procedure were identical to Experiment 1, except that Experiment 2 used two different speakers in each block instead of one single speaker. Between the two speakers, Speaker A (the one used in Experiment 1) was only used in target trials and Speaker B was only used in filler trials. As the base rate of incongruency was manipulated by the filler trials, it was solely associated with Speaker B (the non-target speaker) but not Speaker A (the target speaker). We hypothesized that if a process is specifically related to the target speaker, it should not be affected by the manipulation of base rate which is specifically related to the non-target speaker. Therefore, neurocognitive processes that existed in both Experiment 1 and Experiment 2 should reflect a speaker-general mechanism, while processes that only existed in Experiment 1 but not Experiment 2 should reveal a speaker-specific mechanism.

Experiment 2 included another 30 neurologically healthy native speakers of Mandarin Chinese (15 females, 15 males; mean age = 23.27 years, *SD* = 2.16 years) who were not involved in Experiment 1. The EEG preprocessing excluded 4.64% of trials for amplitude analysis and 5.36% for time-frequency analysis.

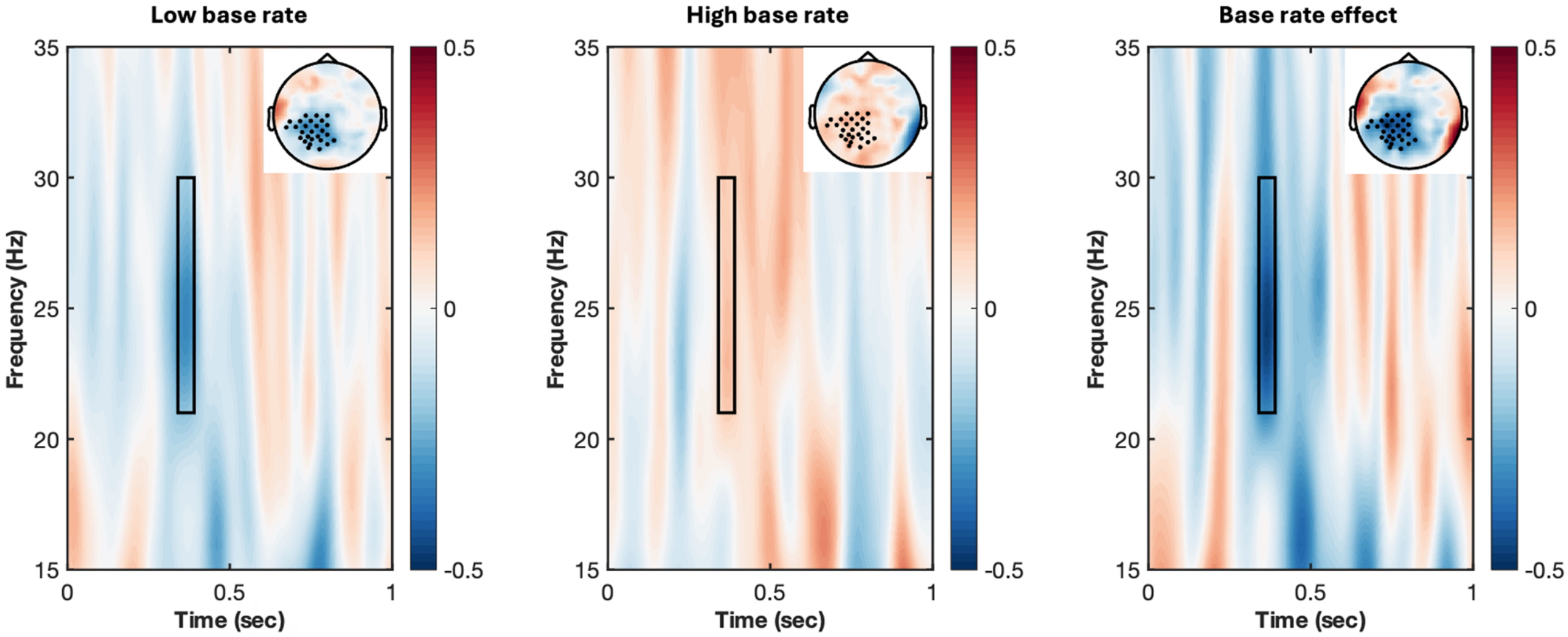

Figure 2. Time-frequency representations of stereotype-incongruency effects of the high-beta (21–30 Hz, 340–390 ms) cluster in Experiment 2.

## Results

Similar to Experiment 1, amplitude analyses showed no significant effects for Congruency, Base rate of incongruency, or their interaction in either the 300–500 ms time window or the 600–1000 ms time window. Models including openness as an additional predictor revealed no significant interactions in either time window (see Table S5 for model structures and full results, see online supplementary material for a colour version of this table).

As shown in Fig. 2, time-frequency analyses identified a significant cluster of high-beta oscillations (340–390 ms, cluster statistic = –301.67, $P = .043$), similar to Experiment 1. LME models on trial-level data (see Table S6 for model structures, see online supplementary material for a colour version of this table) showed a significant interaction between Congruency and Base rate ($\beta = 0.16$, $SE = 0.06$, $t = 2.71$, $P = .007$). In the low base rate condition, stereotype incongruencies led to a significant decrease in high-beta power ($\beta = -0.13$, $SE = 0.04$, $t = -3.16$, $P = .002$), while the incongruency effect was not significant in the high base rate condition ($\beta = 0.02$, $SE = 0.04$, $t = 0.54$, $P = .589$). Further adding openness to the model showed no additional significant effects (see Table S6 for model structures and full results, see online supplementary material for a colour version of this table).

## Discussion

Experiment 2 replicated the high-beta effect observed in Experiment 1, showing that this neural response was modulated by the base rate of stereotype-incongruent statements even when the base rate was associated with a different speaker. However, the theta effect found in Experiment 1 disappeared in Experiment 2. Additionally, we did not observe significant effects in the traditional ERP components, consistent with Experiment 1.

The persistence of the high-beta effect across both experiments suggests that this effect reflects a speaker-general adaptation mechanism, as listeners may adjust their overall expectations about stereotype violations when exposed to frequent stereotype-incongruent statements (even from a different speaker). In contrast, the theta effect's presence in Experiment 1 but absence in Experiment 2 suggests that it is associated with a speaker-specific adaptation mechanism. The disappearance of the theta effect when the base rate manipulation was dissociated from the target speaker indicates that this neural response might specifically relate to the updating of the speaker model. In Experiment 1, listeners could use the statistical properties of the statements to update their mental model of the target speaker, leading to the observed theta modulation. However, in Experiment 2, the base rate information came from a different speaker and thus could not be used to update the mental model of the target speaker, resulting in the absence of the theta effect.

# General discussion

The present study investigated whether listeners adapt their language comprehension based on the probability of speaker-specific statements. Across two experiments, we observed a cluster of high-beta oscillations that were modulated by the general base rate of stereotype-incongruent statements regardless of the specific identity of the speaker. We also identified a cluster of theta oscillations that showed base rate modulation only when it was associated with the target speaker.

We did not observe ERP effects in amplitude analyses. This absence might stem from our experimental design. Previous studies on speaker-stereotype incongruency (e.g. Van Berkum et al. 2008; Wu and Cai 2025) often included multiple speakers and thus created a context that encouraged listeners to simply integrate the speaker demographics with their population-level stereotypes (regardless of the specific identity of the speaker). This type of integration is well-indexed by ERP components like the N400. In contrast, our focus on a single, repeating speaker (i.e. the target speaker) likely promoted a shift from this general integration to an active, dynamic process of building and updating a mental model for that specific individual. This active modelling process might be better reflected by the non-phase-locked oscillatory dynamics we observed.

The high-beta oscillations showed sensitivity to the base rate of stereotype-incongruent statements regardless of specific speaker identity, suggesting a speaker-general adaptation mechanism. The differential modulation of high-beta power by stereotype incongruency in the low and high base rate conditions may relate to the maintenance versus change of cognitive states (Engel and Fries 2010). Increases in beta power might signal the maintenance of the current cognitive state, while decreases in beta power could indicate state changes. In the low base rate condition, where stereotype-incongruent statements were rare and unexpected, the significant decrease in high-beta power likely reflected the comprehension system's detection that the current processing state needed to change to handle the unexpected input. Conversely, in the high base rate condition, where stereotype-incongruent statements were frequent and thus expected, the increase in high-beta power suggests the system was actively maintaining a processing state adapted to handle stereotype incongruency. This interpretation aligns with proposals regarding beta oscillations' role in maintaining versus changing cognitive states during language comprehension (Lewis et al. 2015).

The theta effect observed only in Experiment 1 appears to reflect a speaker-specific adaptation mechanism. The direction of this effect was modulated by the base rate of incongruency: In the low base rate condition, where stereotype-incongruent statements were rare and unexpected, the decrease in theta power might reflect the deployment of attention to those statements, as it is often associated with a generalized decrease in low-frequency power that is thought to reflect an upregulation of mechanisms that down-weight noise and up-weight information with high task relevance in the inputs (Herweg et al. 2020). Conversely, in the high base rate condition, where stereotype-incongruent statements were frequent and thus expected, the increase in theta power suggests active maintenance of the speaker-specific representations in working memory (Khader et al. 2010).

This speaker-specific adaptation was also modulated by listeners' openness trait. For individuals of high openness, the tendency to show a theta power decrease in response to incongruency is consistent with a greater deployment of attention to these stereotype-violating events. This may reflect a flexible cognitive style where each piece of incongruent information is treated as salient and worthy of attention for model updating. Conversely, the tendency for individuals of low openness to show a theta power increase suggests that they may engage in more effortful maintenance of their initial stereotype-based speaker model, with the theta increase likely reflecting the increased cognitive and memory load required to manage this conflicting information.

Our interpretation of the selective influence of openness on theta but not high beta power is that openness, a personal trait reflecting flexibility in revising one's beliefs, would be most relevant to the process of updating a mental model about a specific speaker. The speaker-general beta effect occurred relatively early (Experiment 1: 220–330 ms; Experiment 2: 340–390 ms), consistent with a more lower-level adjustment to environmental statistics. In contrast, the speaker-specific theta effect emerged later (320–580 ms), aligning with the time course of more controlled, higher-level updating processes. It is plausible that these more deliberative cognitive operations are more susceptible to the influence of a high-level personality trait like openness.

It should be noted, however, that our interpretation of theta oscillations is based on indirect evidence: the presence of theta modulation in Experiment 1, where the speaker's own statistics were manipulated, and its absence in Experiment 2, where another speaker's statistics were manipulated. More direct evidence would be a demonstration of this specificity within a single experiment. For example, future research could manipulate the statement statistics of two different speakers independently within the same block in order to further test whether theta oscillations for a given speaker are modulated exclusively by that speaker's own base rate of incongruency.

These findings extend previous work on probabilistic adaptation in language comprehension. Gibson et al. (2013) demonstrated that comprehenders rationally adapt to the base rate of implausible sentences during syntactic processing. The present study shows that similar probabilistic adaptation mechanisms are also involved in speaker-contextualized language processing. Just as comprehenders track and adapt to the probability of certain linguistic structures, they also track and adapt to the probability of a speaker making a certain kind of statements.

In conclusion, we show that language comprehension is probabilistically adapted for individual speakers. Rather than exclusively relying on fixed prior knowledge, listeners construct and dynamically refine probabilistic speaker models in light of the linguistic and speaker information they perceive. These findings extend the speaker model account by showing how such models are updated through experience and provide evidence for the interaction between social cognition and language processing.

## Acknowledgements

The authors thank Siyuan Zhao and Hongyan Wang for their assistance during data collection.

## Author contributions

Hanlin Wu (Conceptualization [equal], Data curation [equal], Formal analysis [equal], Investigation [equal], Methodology [equal], Project administration [equal], Resources [equal], Software [equal], Visualization [equal], Writing—original draft [equal], Writing—review & editing [equal]), Xiaohui Rao (Investigation [equal], Methodology [equal], Project administration [equal], Resources [equal]), and Zhenguang G. Cai (Conceptualization [equal], Funding acquisition [equal], Supervision [equal], Writing—original draft [equal], Writing—review & editing [equal])

## Supplementary data

Supplementary data is available at *SCAN* online.

Conflict of interest: None declared.

## Funding

This work was supported by the General Research Fund to Zhenguang G. Cai (grant number: 14600220), University Grants Committee, Hong Kong.

## Data availability

The stimuli and data set of this study are available at https://osf.io/zet3r/.

## Ethical research

The study protocol was under the ethical standards of the Helsinki Declaration and approved by the Joint Chinese University of Hong Kong-New Territories East Cluster Clinical Research Ethics Committee.